\documentclass[aps,pre,onecolumn,superscriptaddress,showpacs,floatfix,longbibliography]{revtex4-1}
\usepackage{bm}
\usepackage{amssymb}
\usepackage{epsfig}
\usepackage{amsmath}
\usepackage{color}
\usepackage{graphicx}
\usepackage{epstopdf}
\usepackage[toc]{appendix}

\begin{document}

\title{Detecting and characterizing high frequency oscillations in epilepsy - A case study of big data analysis}

\author{Liang Huang}
\affiliation{School of Physical Science and Technology, Lanzhou University, Lanzhou, Gansu 730000, China}

\author{Xuan Ni}
\affiliation{School of Electrical, Computer, and Energy Engineering,
Arizona State University, Tempe, AZ 85287, USA}

\author{William L. Ditto}
\affiliation{College of Sciences, North Carolina State University, Raleigh, NC 27695, USA}

\author{Mark Spano}
\affiliation{School of Biological and Health Systems Engineering, Arizona
State University, Tempe, AZ 85287-9709, USA}

\author{Paul R. Carney}
\affiliation{Pediatric Neurology and Epilepsy, Department of Neurology,
University of North Carolina, 170 Manning Drive, Chapel Hill,
NC 27599-7025, USA}

\author{Ying-Cheng Lai} \email{Ying-Cheng.Lai@asu.edu}
\affiliation{School of Electrical, Computer, and Energy Engineering,
Arizona State University, Tempe, AZ 85287, USA}
\affiliation{Department of Physics, Arizona State University, Tempe,
AZ 85287, USA}


\date{\today}

\begin{abstract}
We develop a framework to uncover and analyze dynamical anomalies from
massive, nonlinear and non-stationary time series data. The framework consists
of three steps: preprocessing of massive data sets to eliminate erroneous
data segments, application of the empirical mode decomposition and Hilbert
transform paradigm to obtain the fundamental components embedded in the
time series at distinct time scales, and statistical/scaling analysis of
the components. As a case study, we apply our framework to detecting and
characterizing high frequency oscillations (HFOs) from a big database
of rat EEG recordings. We find a striking phenomenon: HFOs exhibit
on-off intermittency that can be quantified by algebraic scaling laws.
Our framework can be generalized to big data-related problems in other
fields such as large-scale sensor data and seismic data analysis.

\end{abstract}

\maketitle

\section{Introduction}

Big data analysis~\cite{Marx:2013,SS:2013,KWG:2013,CMZL:2014,FHL:2014,
LKKV:2014}, a frontier field in science and engineering, has broad
applications ranging from biomedicine and smart health~\cite{Howeetal:2008,
BG:2013} to social behavior quantification and energy optimization in civil
infrastructures. For example, in biomedicine, vast electroencephalogram
(EEG) or electrocorticogram (ECoG) data are available for the analysis,
detection, and possibly prediction of epileptic seizures (e.g.,
Refs.~\cite{SMNPDC:2006,THSSFMWDC:2008,KRTSKSFDC:2009,CMDDTHC:2009,
THDSC:2009,FTCD:2010,NTKCD:2010,CDMPTHDDC:2010}). In a modern
infrastructure viewed as a complex dynamical system, large
scale sensor networks can be deployed to measure a number of physical
signals to monitor the behaviors of the system in continuous
time~\cite{TYWLW:2012,GGA:2013,AZWSZ:2013}. In a modern city, smart
cameras are placed in every main street to monitor the traffic flow
at all time. In a community, data collected from a large number of
users carrying various mobile and networked devices can be used for
community activity prediction~\cite{ZCMH:2014}. In wireless
communication, big data sets are ubiquitous~\cite{CML:2014,SM:2014}. In
all these cases, monitoring, sensing, or measurements typically result
in big data sets, and it is of considerable interest to detect behaviors
that deviate from the norm or the expected.

In this paper, we develop a general and systematic framework to
detect hidden and anomalous dynamical events, or simply anomalies, from
big data sets. The mathematical foundation of our framework is
Hilbert transform and instantaneous frequency analysis. The reason for
this choice lies in the fact that complex dynamical systems are
typically nonlinear and non-stationary. 
For such systems, the traditional Fourier analysis is limited 
because, fundamentally, they are designed for linear and stationary 
systems. Windowed Fourier analysis may be feasible to generate 
patterns in the two-dimensional frequency-time plane pertinent to 
characteristic events, but two-dimensional feature identification 
is difficult. In contrast, the features generated by the EMD methodology 
are one-dimensional, which are easier to be identified computationally.
The Hilbert transform and instantaneous frequency-based analysis have proven to
be especially suited for data from complex, nonlinear, and non-stationary
dynamical systems~\cite{Huangetal:1998,YL:1997,QKKG:2002}.
The challenge is to develop a mathematically justified and computationally 
reasonable framework to uncover and characterize ``unusual'' dynamical
indicators that may potentially be precursors to a large scale, catastrophic 
dynamical event of the system.

The general principle underlying the development of our big data-based
detection framework is as follows. First, we develop an efficient
procedure for pre-processing big data sets to
exclude erroneous data segments and statistical outliers. Next, we
exploit a method based on a separation of time scales, the empirical
mode decomposition (EMD) method~\cite{Huangetal:1998,YL:1997}, to detect
anomalous dynamical features of the system. Due to
its built-in ability to obtain from a complex, seemingly random time series
a number of dominant components with distinct time scales, the method is
anticipated to be especially effective for anomaly detection. We pay
particular attention to the challenges associated with big data sets.
Finally, we perform statistical analysis to identify and characterize
the anomalies and articulate their implications.

As a concrete example to illustrate the general principle of our big data
analysis framework, we address the detection of high frequency oscillations
(HFOs), which are local oscillatory field potentials of frequencies greater
than 100 Hz and usually have a duration less than one
second~\cite{SWBFE:2002,WPCJBL:2004,JULODG:2006,WGSH:2008,ZJZDG:2009,
BES:2010,CNHC:2010,MZV:2011,Blancoetal:2011,ZJZLJG:2012,Jacobsetal:2012,
Haegelenetal:2013}. Oscillations between 100 and 200 Hz are called ripples
and occur most frequently during episodes of awake immobility and slow wave
sleep. The HFOs in this range are believed to play an important role in
information processing and consolidation of memory~\cite{Buzsaki:1996,
SW:1998}. Pathologic HFOs (with frequency larger than 200 Hz, or fast
ripples~\cite{BWSRFE:2002}) reflect fields of hyper-synchronized action
potentials within small discrete neuronal clusters responsible for
seizure generation. They can be recorded in association with
interictal spikes only in areas capable of generating recurrent
spontaneous seizures~\cite{BEW:1999}. Thus detecting fast ripple can
be useful in locating the seizure onset zone in the epileptic
network~\cite{UCDG:2007,WGSH:2008,EBSM:2009}, and this was verified
previously using data sets from a wide variety of patients~\cite{CNHC:2010}.
In particular, it was found that almost all fast-ripple HFOs were recorded
in seizure-generating structures of patients suffering from medial or
polar temporal-lobe epilepsy, indicating that the ripples are a specific,
intrinsic property of seizure-generating networks in these brain areas.
The pathologic HFOs and their spatial extent can potentially be used as
biomarkers of the seizure onset zone, facilitating decisions as to whether
surgical treatment would be necessary. Besides their role in locating the
seizure onset zone, HFOs may also reflect the primary neuronal disturbances
responsible for epilepsy and provide insights into the fundamental
mechanisms of epileptogenesis and epileptogenicity~\cite{KQGB:2005,QKB:2006}.

Traditional methods such as the Fourier transform and spectral analysis
assume stationarity and/or approximate the physical phenomena with
linear models. These approximations may lead to spurious components in
their time-frequency distribution diagrams if the underlying signal is
non-stationary and nonlinear. Empirical Mode Decomposition (EMD) is a
technique~\cite{Huangetal:1998} to specifically deal with non-stationary
and nonlinear signals. Given such a signal, EMD decomposes it into
a small number of modes, the intrinsic mode functions (IMFs), each having a
distinct time or frequency scale and preserving the amplitude of the
oscillations in the frequency range. The decomposed modes are orthogonal
to each other, and the sum of all modes gives the original data. The ease
and accuracy with which the EMD method processes non-stationary and
nonlinear signals have led to its widespread use in various applications
such as seismic data analysis~\cite{Huangetal:1998}, chaotic time series
analysis~\cite{YL:1997,Lai:1998}, neural signal processing in biomedical
sciences~\cite{LLC:2005}, meteorological data analysis~\cite{Duffy:2005},
and image analysis~\cite{NBDNB:2003}. We develop an EMD
based method to detect HFOs. Due to its built-in
ability to pick out from a complex, seemingly random time series a number
of dominant components of distinct time scales, the method is
especially effective for the detection of HFOs. We finally perform
a statistical analysis and find a striking phenomenon: HFOs occur in
an on-off intermittent manner with algebraic scaling. In addition to
HFOs, our framework can detect population spikes, oscillations in the
frequency range from 10 to 50 Hz, as well as distinct and independent IMFs.

Since pathologic HFOs reveal dynamical coherence within small discrete
neuronal clusters responsible for seizure generation, a good understanding
and accurate detection of HFOs may bring the grand goal of early seizure
prediction one step closer to reality and would also improve the
localization of the seizure onset zone to facilitate decision making
with regard to surgical treatment. Not only does our method illustrate,
in a detailed and concrete way, an effective way to analyze big data
sets, our finding also has potential impact in biomedicine and human health.

There were existing works on applying the EMD/Hilbert transform method 
to neural systems. Earlier the method was applied to analyzing biological 
signals and performing curve fitting~\cite{Huang:2004}, and a combination
of EMD, Hilbert transform, and smoothed nonlinear energy operator was 
proposed to detect spikes hidden in human EEG data~\cite{CLOG:2005}.  
Subsequently, it was demonstrated~\cite{Li:2006} that the methodology can be 
used to analyze neuronal oscillations in the hippocampus of epileptic rats in 
vivo with the result that the oscillations are characteristically different 
during the pre-ictal, seizure onset and ictal periods of the epileptic EEG 
in different frequency bands. In another work~\cite{SN:2007}, 
the EMD/Hilbert transform method was applied to detecting synchrony 
episodes in both time and frequency domains. The method was demonstrated to be 
useful for decomposing neuronal population oscillations to gain insights 
into epileptic seizures~\cite{LJFY:2008}, and EMD was used for extracting
single-trial cortical beta oscillatory activities in EEG 
signals~\cite{YCWL:2010}. The outputs of EMD, i.e., the IMFs, were 
demonstrated to be useful for EEG signal classification~\cite{BP:2012}. Our 
work differs from these previous works in that we address the issue of 
detecting HFOs and uncovering the underlying scaling law.

\section{Results} \label{sec:results}

\subsection{Pretreatment of data sets}

High sampling (12KHz), multichannel (32-64 channels), continuous
recordings of local field potentials in freely moving rodents presents
unique technical challenges. Although most channels continue to record
over a 4-6 week periods, over time the integrity of the signal degrades
and electrode recording may come off and on line. To this end, it is
important to pre-process data files to exclude gaps in data. This in
itself is challenging due to the large size of each dataset (about 5 Terabytes),
variability during recordings of local field potentials, and gaps in data.
Here, we develop a fully automated statistical method. The resulting
``data-mining'' algorithm is general and we expect it to be useful for
dealing with other massive data sets.


For our study we examine EEG data taken from a rat model of the approach
to epilepsy. The typical size of a binary file in our database is about
$600-700$ MB. Each file belongs to a certain channel (specified by a channel
number) and a specific time duration (specified by a file number). We regard
the channel and file numbers as two orthogonal dimensions and plot the
contour of a suitable statistical quantity (to be discussed below) in the
two dimensional plane, so the data of one rat ($\sim4$ Terabytes) can be
represented by a single contour plot. The whole process can be
programmed to be highly parallelized, providing a global overview
of the raw EEG data.

Let $d_i$, $i=1,\dots,L$ be the value of the EEG signal for a single sample,
where $L$ is the number of samples in a binary file. In the experiment, each
value $d_i$ is recorded as a $16$-bit integer, so
$d_i\in[-N,N-1]$, where $N = 2^{15}$ and, typically we have
$L\sim3\times10^8$ samples. We then examine the values of
$d_i$ and count the number of each value present in the file,
which results in an array $n_j$, $j=-N,\dots,N$. Repeated values over
some periods in the oscillation pattern lead to corrupted files,
which can be due to recording errors - indeed happened in our experiment. 
In general, when hours or even minutes of bad recordings of zeros are
encountered, the number $n_0$ of zeros in the file counted will
increase rapidly. These features of abnormal recordings will be 
utilized to exclude the corrupted files.

\begin{figure}
\centering
\includegraphics[width=\linewidth]{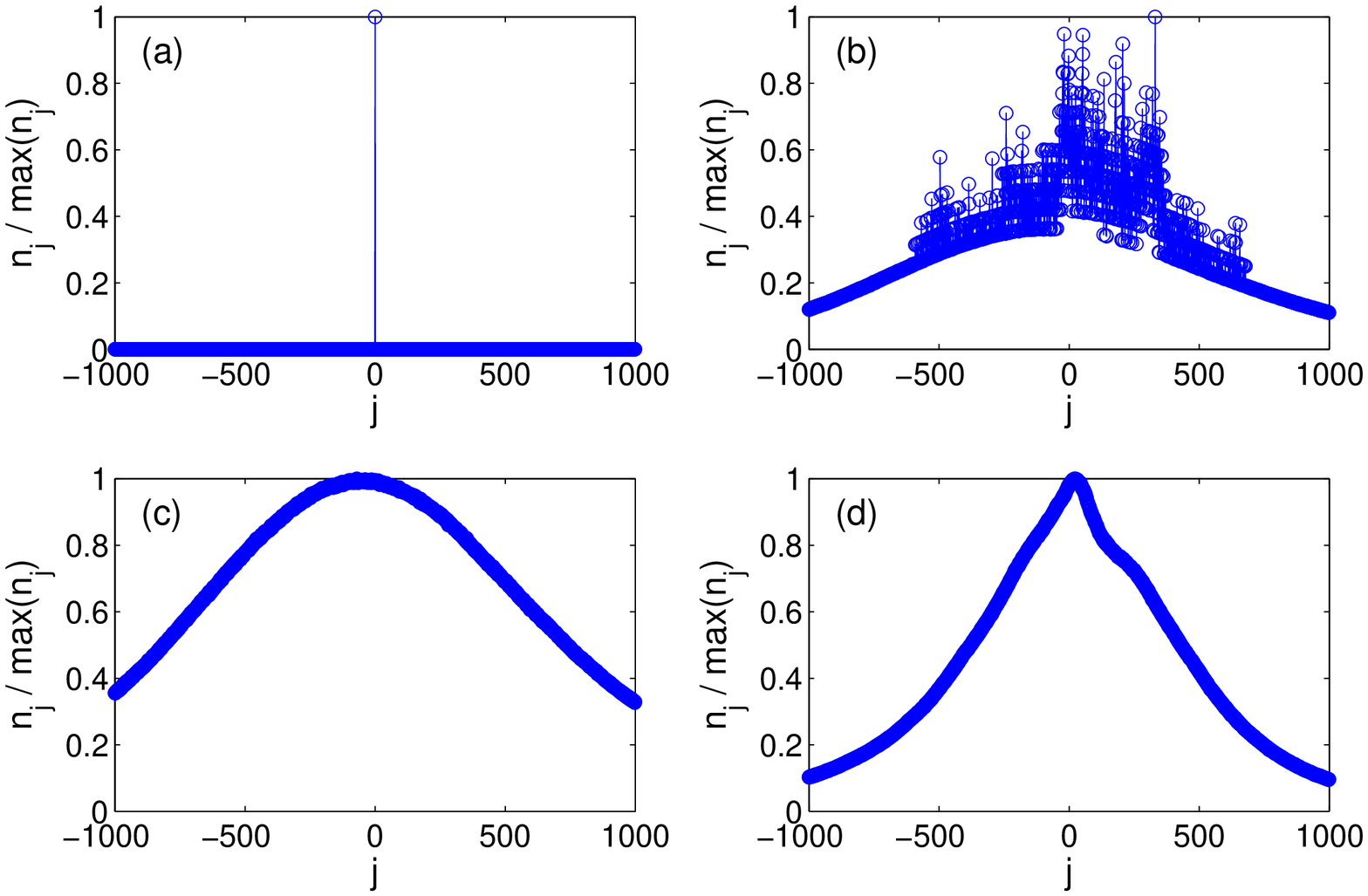}
\caption{\textbf{Pretreatment of massive data from rat EEG.}
Different types of distributions for Rat004 channel 02. For each panel,
the $y$-axis is normalized by the maximum of $n_j$. The four panels
correspond to: (a) a corrupted file with a large number of zeros
(file $\#$20), (b) a bad recording with repetitions of oscillating
patterns (file $\#$73), (c) a normal files without transitions
(file $\#$77), and (d) a file containing a seizure (file $\#$99).}
\label{fig:stattypes}
\end{figure}

Figure \ref{fig:stattypes} shows four typical types of $n_j$
distributions obtained from channel $02$ of Rat004 composed of $229$
data files. For binary files that have large numbers of continuous
zeros, for example, file number $20$, the distribution is
shown in Fig.  \ref{fig:stattypes}(a). An example of a corrupted
file is where a specific pattern of oscillations is embedded
repeatedly in most of the data in the file. The corresponding
distribution is shown in Fig. \ref{fig:stattypes}(b). The distributions
from normal data files qualified for dynamical analysis are shown in
Figs. \ref{fig:stattypes}(c) and \ref{fig:stattypes}(d). The distribution
in Fig. \ref{fig:stattypes}(c) is
approximately Gaussian. However, seizure events can cause distortions
from the Gaussian distribution, as evidenced by Fig. \ref{fig:stattypes}(d)
for file number $99$ where a clinically certified seizure
is present. We observe that the distribution becomes somewhat narrowed
(as compared with the case of no seizure) and slightly asymmetrical.

After obtaining the distribution for each file, we define and
compute a statistical quantity for each file, and assemble the files
within the same channel according to this quantity, as follows.
Let $s_k=n_{k-N}-n_{k-N-1}$ ($k=1,\dots,2N$), where $s_k$ represents
the difference between two neighboring counts, and let
$\sigma_s^2 = 1/(2N-1)\sum_{k=1}^{2N}(s_k-\bar{s})^2$ be the variance,
where $\bar{s}$ is the mean value of $s_k$. 
Note that $n_j$ is not normalized and their sum is the data length $L$
of the file. Denoting $\langle n\rangle_j$ as the smoothed curve of $n_j$,
we have $\langle n\rangle_j \sim L$. The fluctuations, as characterized
by $s_k$, are in general proportional to $\langle n\rangle_j$. As a result,
the variance $\sigma_s^2$ is positively correlated with $L$, e.g., a
larger value of $L$ would result in a larger value of $\sigma_s^2$.

\begin{figure}
\centering
\includegraphics[width=\linewidth]{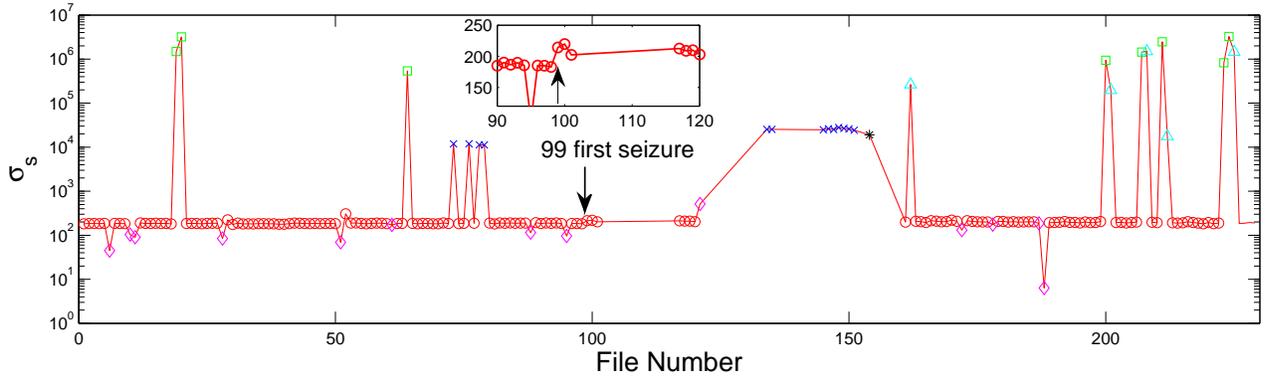}
\caption{\textbf{Statistical properties of massive data from rat EEG.}
Standard deviation $\sigma_s$ for Rat004 channel 02. Red circles denote
the normal files; green squares are the files with large numbers of
zeros; blue crosses are corrupted files; pink diamonds are small
files; cyan triangles are small files with many zeros; black star
is small corrupted files. The arrow marks the file $99$ which has
the first seizure. The inset shows the enlarged area around file
$99$ on a linear scale.}
\label{fig:logsigma}
\end{figure}

Thus for small files that are normal in all other aspects, we will
have smaller $\sigma_s$ values, which can be clearly identified from Fig.
\ref{fig:logsigma} as the points below the majority. This figure
also shows some points with extremely large $\sigma_s$ values. These
points correspond to the binary files with large numbers of
continuous zeros [Fig. \ref{fig:stattypes}(a)]. The corrupted data
all have $\sigma_s$ values in the range $10^4\sim10^5$
[Fig.~\ref{fig:stattypes}(b)], which can be excluded readily,
as shown in the inset of Fig.~\ref{fig:logsigma}. A transition in
$\sigma_s$ at file number 99 (when the first seizure occurred) is observed.

\begin{figure}
\centering
\includegraphics[width=\linewidth]{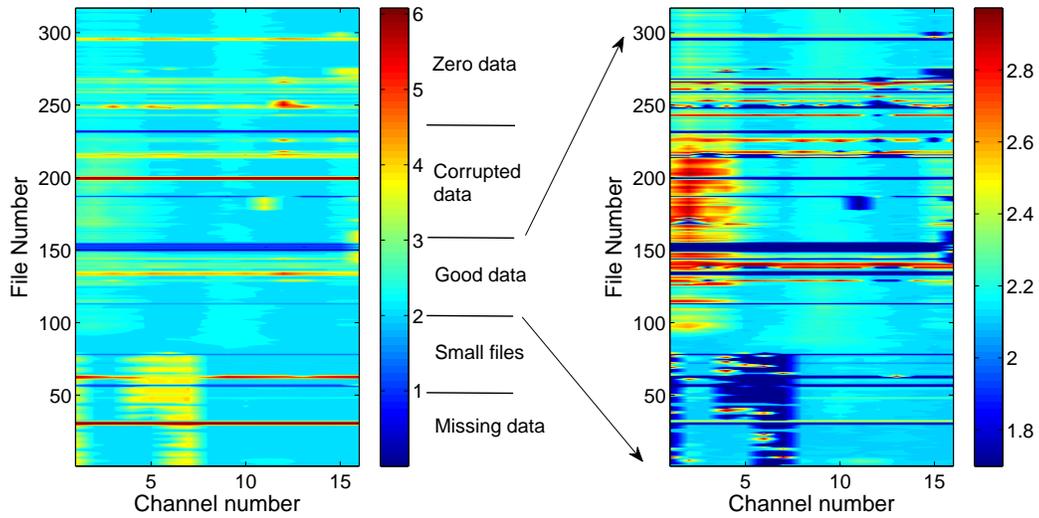}
\caption{\textbf{Contour representation of massive data from rat EEG.}
Contour plot of $\log_{10}\sigma_s$ for Rat001. Left
panel shows the whole $\sigma_s$ range. Different types of data are
classified according to the value of $\log_{10}\sigma_s$. In the
right panel, the contour is for values of $10^2\le\sigma_s\le10^3$
(good data), and the remaining values of $\sigma_s$ are set to $50$ so
that the dark blue area marks all abnormal data.}
\label{fig:contour}
\end{figure}

By applying the same procedure to multiple channels, the massive EEG data
from one rat can be expressed using a single contour plot of $\sigma_s$,
as shown in Fig.~\ref{fig:contour}. We can immediately identify different
types of data in terms of the values of $\log_{10}\sigma_s$. 
It should be noted that for different ``abnormal'' situations, e.g., small 
files, contamination with zeros, or corrupted data, there can be different 
methods of remedy based on examining different aspects particular to the 
data. However, our method is general and efficient in that a single 
indicator is effective at distinguishing the different types of 
abnormalities in the files during the pre-processing stage of the 
massive database for further dynamical and statistical analysis.

\subsection{EMD analysis of EEG data}

We conducted extensive tests of applying the EMD procedure to EEG data
(see {\bf Methods}). 
A general finding is that the resulting IMFs in different frequency ranges 
possess statistical features that are relevant to certain brain activities, 
demonstrating that the EMD methodology can be effective for probing the 
dynamical origins of epileptogenesis. For example, typically
the frequencies of the first 5 IMFs are about 5 kHz, 2 kHz, 1 kHz, 500 Hz,
200 Hz, 100Hz. Since the sampling frequency is 12 kHz, the first three
modes correspond to mostly noise contained in the original EEG data.
The fourth to sixth modes, whose frequencies lie in the range between
50 Hz and 800 Hz represent the intrinsic dynamical evolution of the
underlying brain system.

Our procedure for analyzing long EEG data thus consists of
performing the EMDs to obtain different IMFs, calculating
the amplitudes and frequencies of the IMFs that are deemed to
reveal the dynamical evolution of the brain, and
performing statistical analysis of the on-intervals for the 
IMFs in a proper frequency range.
An example is shown in Fig.~\ref{fig:distribution}, where the distributions 
of the amplitude and frequency of an IMF for a particular channel of 
two-months' duration are shown. The entire data set was divided
into 230 files, each containing 7 hours of EEG recording. 
Note that when performing the EMD, the data is broken into small segments,
e.g., 5 second for each segment, to make the computation more efficient. 
To calculate the IMFs, a 0.5 second segment is included at each end of   
the 5-second segment to eliminate the edge effect so that the IMFs can
be accurately determined. From Fig.~\ref{fig:distribution}, we can 
distinguish the changes in the rat brain activity, such as stimulation and the
occurrence of the first seizure. Recurrent seizures are not so clearly
visible in this plot. Another apparent feature revealed is the
circadian periodicity. The EEG recording has a 24 hour periodicity
because of the circadian activity or of the external treatment of the
rat such as feeding, etc., which also changes the
frequency and amplitude of each decomposed IMF. Since each file
is 7 hours long, the circadian periodicity indicates a periodicity of 3
(or 4) files in the plot, which is apparent from the comb-like
structure in the plot, especially in Fig. \ref{fig:distribution}(a),
where adjacent two comb teeth have a separation of about 3 files.

\subsection{Detection of HFOs and population spikes}

\begin{figure}
\centering
\includegraphics[width=\linewidth]{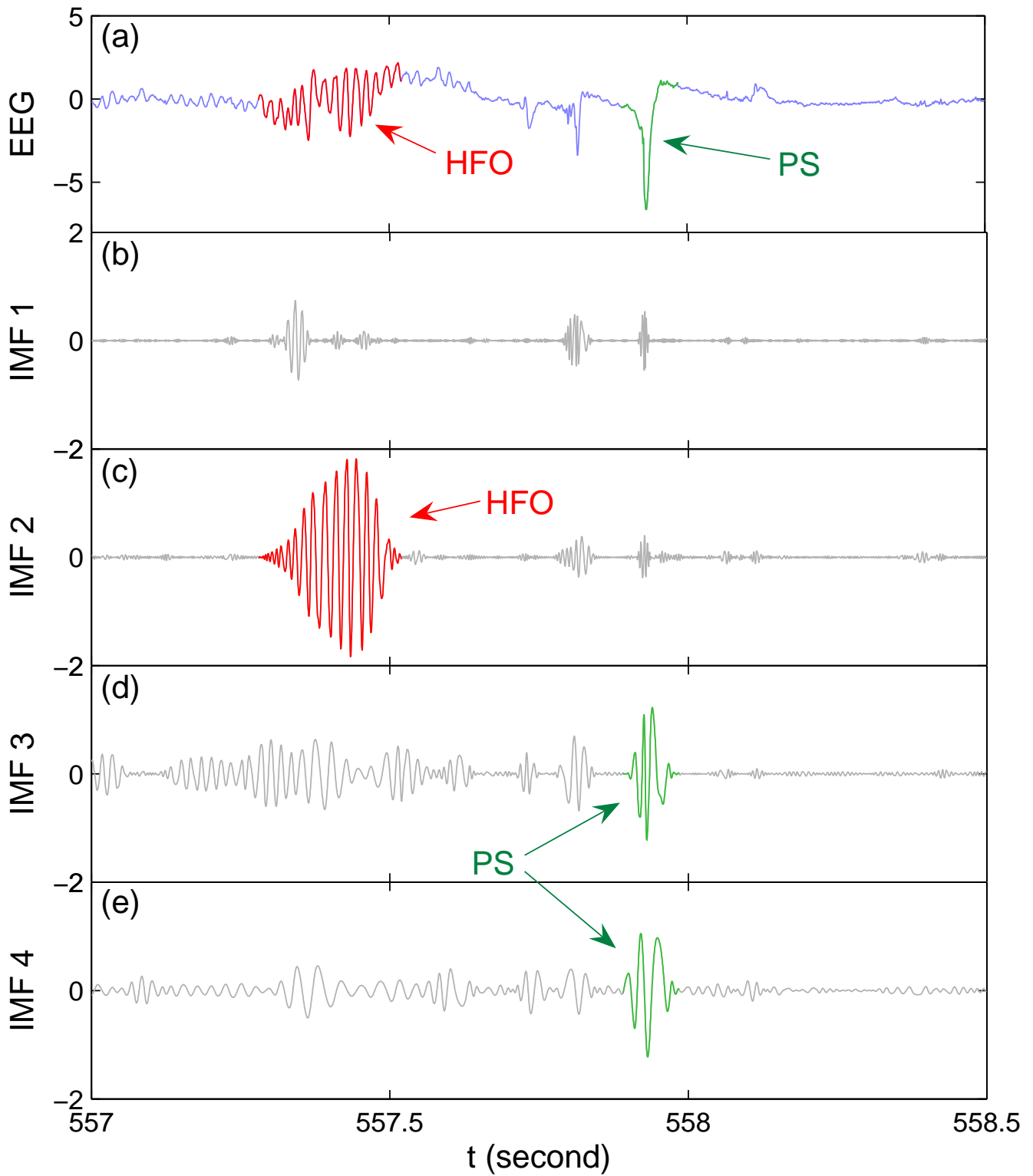}
\caption{\textbf{Example of EMD-based HFO detection from EEG data.}
A $1.5$ second segment of normalized EEG data containing an HFO and a
population spike. (b)-(e) are the IMFs in the frequency range of interest.
The HFO is revealed in IMF 2 and the population spike is revealed in IMF 3
and IMF 4.}
\label{fig:HFOPS}
\end{figure}

To illustrate the detection of HFOs and population spikes, we reduce the
sampling rate so that these dynamical events can be visualized clearly.
Note that, when the sampling rate is reduced, the noisy components are
effectively filtered out, so the first few IMFs become important. (In
the detection of HFOs and population spikes, higher sampling
frequency should still be used, in which case the first few IMFs need to
be disregarded, as discussed above.) Figure~\ref{fig:HFOPS} shows, for a
segment of down-sampled EEG data, the relevant empirical modes. For this
data set, the frequencies are 200-500 Hz for mode 1, 80-200 Hz for mode 2,
about 50 Hz for mode 3 and 30 Hz for mode 4. Take mode 2 as an example,
the IMF will have small amplitude if the original EEG data does not
contain oscillations in the corresponding frequency range. When the EEG
data contains these oscillations, they will be revealed in the corresponding
IMF. Since HFOs are generally associated with frequencies larger than
80 Hz, they will be revealed in the first two modes. The population
spikes (with a time scale of 0.1 s) are decomposed by EMD into
oscillations in the frequency range 10-50 Hz, thus they will mainly be
manifested in modes 3 and 4. The EEG data in Fig.~\ref{fig:HFOPS}(a)
contains an HFO and a population spike. It is apparent that the HFO and
population spike are separated by EMD into different modes and are
localized in different time scales, e.g., Fig.~\ref{fig:HFOPS}(c) for the
HFO and Figs.~\ref{fig:HFOPS}(d,e) for the population spike. Thus the
amplitude of the modes evolving in time can be used to detect the HFOs or
population spikes, depending on the frequency range of the mode.

Our results thus suggest strongly the feasibility of developing
EMD based algorithms to systematically detect all the HFOs
and population spikes. In this regard, we note an existing method of
detection of HFOs, which employs short-time energy or line
length of the acquired data for HFOs in some small frequency
ranges~\cite{GWMDL:2007}. Our method is capable of detecting HFO and can be used to distinguish various
oscillation profiles. Based on the detected HFOs and population spikes,
extensive statistical analyses for the five critical phases during
epileptogenesis, namely, pre-stimulation state, pre-seizure state, status
epilepticus phase, epilepsy latent period, spontaneous/recurrent seizure
period, can be carried out to gain {\em unprecedented} insights into
epileptogenesis.

It can occur that, for a particular signal whose highest frequency
component is most significant to the underling dynamics, the first
IMF contains the dominant dynamics with the highest frequency, the next
is a lower frequency background to it, and so on. However, while the
first IMF is the highest frequency component, the corresponding frequency
range may not necessarily be relevant to the system dynamics. In fact,
we found that typically the first IMF corresponds to noise, and the next
IMF contains information about the dynamics of the system.
Which IMFs are actually useful and informative depends on the nature
of the original signal. More specifically, what EMD does is to decompose
the signal into different frequency components through different IMFs, which
contain the time varying amplitude and frequency information for each
component embedded in the original signal. If the signal is contaminated
by noise, the first IMF would be the noise component that contains
little information about the underlying dynamics. Our analysis of the massive
EEG data indicates that this is indeed the case.

\subsection{Automated detection and classification of HFOs}

\begin{figure}
\centering
\includegraphics[width=\linewidth]{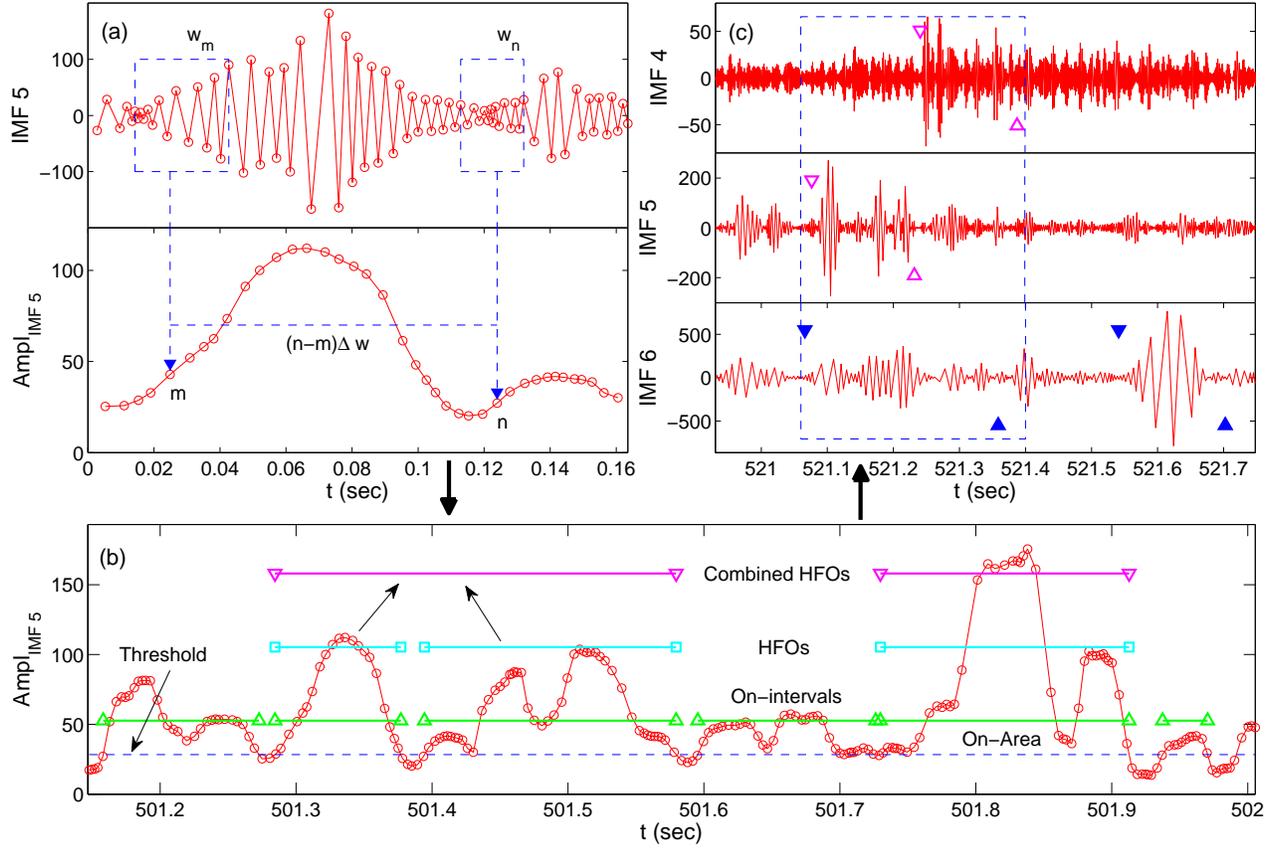}
\caption{\textbf{Illustration of HFO detection method.}
The method consists of three steps.
(a) Computing the amplitude function from each IMF generated by
EMD. The size of the moving window is $w = 7$ periods
(indicated by the blue dashed boxes). The time step for
the moving window is $\Delta w$.
(b) For each IMF, we locate the on-intervals, find HFOs, and combine
adjacent HFOs if they are too close to each other. The blue dashed
line is the threshold chosen for the segment of the amplitude
function. (c) Classifying HFOs in terms of their frequencies, e.g.,
ripples (solid blue triangles), fast-ripples (open magenta
triangles), and then combining overlapping HFOs across different IMFs,
as shown in the blue dashed box.}
\label{fig:scheme}
\end{figure}

Our method to detect, characterize, and understand HFOs from
EEG recordings consist of the following steps: (1) performing EMD and
calculating distinct IMFs, (2) searching for HFOs based on the amplitudes of
IMFs, and (3) classifying HFOs in terms of their frequencies and
calculating the statistical properties of HFOs. An illustration
of the steps is shown in Fig.~\ref{fig:scheme}.

We first calculate the amplitudes of the IMFs from an automated EMD
procedure. We then locate the extrema of one IMF and define the
interval between two neighboring maxima (or minima) to be one period $T$,
as shown in Fig. \ref{fig:scheme}(a). Unlike the Fourier transform
that becomes ineffective in time-series analysis when the signal
frequency changes with time, EMD is well suited for
generating IMFs whose frequencies vary with time, i.e., when
the period is a function of time: $T=T(t)$. We set a moving time window
of size $w$ and calculate the average IMF amplitude within the window.
The window contains a fixed number of IMF periods. Since the period
varies with time, the actual time span of the window also changes
with time. As an example, we show in Fig.~\ref{fig:scheme}(a)
two windows $w_m$ and $w_n$, each containing the same number (7) of
IMF periods. Apparently, their sizes are different. 
The amplitude values are weighted magnitudes and are calculated as 
$A_m=(1/2)\sum_{i\in w_m}(x_{i+1}-x_i)(|y_{i+1}|+|y_i|)$, where $x_i$
and $y_i$ are the position and magnitude of the IMF, respectively, and 
the factor $(x_{i+1}-x_i)$ is the corresponding weight.
The calculation is repeated when the window is moved to the next position 
by the step size $\Delta w$, which is also chosen to contain a certain 
number of IMF periods. 
Small values of $w$ and $\Delta w$ result in rapidly
oscillating amplitude functions, whereas too large values of $w$ and
$\Delta w$ would cause a loss of information. Empirically, the window 
size can be chosen to include several IMF periods, e.g., six to nine, 
where the moving forward step size $\Delta w$ can be set as one.

The next step is to find on-intervals, time intervals when the amplitude
values are larger than a certain threshold $A_c$, as shown in
Fig.~\ref{fig:scheme}(b). To set a proper threshold, we can pick a
segment (e.g., one hour long) from the amplitude data and calculate the
mean $\mu$ as well as the standard deviation $\sigma$. One way to set the
threshold is $A_c=a_\mu\mu + a_\sigma\sigma$, where $a_\mu$ and $a_\sigma$
are two adjustable parameters on the order unity. To characterize each
on-interval $O_i$, we define an on-area $S_i$, the area of the IMF in the
on-interval above the threshold:
\begin{equation}
S_i=(1/2)\sum_{j\in O_i}(x_{j+1}-x_j)(A_{j+1}+A_j-2A_c),
\end{equation}
where $x_j$ and $A_j$ are the position and magnitude of the amplitude
function of the underlying IMF, respectively. 
The notation $S_i$ with the capital $S$ should be distinguished from $s_k$ 
that denotes the difference between two neighboring counts.
On-intervals are sorted in
the descending order in terms of their on-areas: $S_1>S_2>\cdots$. The
HFOs are identified as those with the largest on-areas, i.e., with
longer duration and larger oscillating amplitudes. Since the on-area
values of HFOs are typically much larger than those of non-HFO
on-intervals, the HFOs can be reliably identified as the outliers in the
on-area statistics, as follows.

Starting from the most significant on-interval $O_1$, we can evaluate its
on-area $S_1$ with a score function of the remaining on-intervals. If
\begin{equation}
S_1 > \alpha E[\{S_i\}_{i\ge2}] + \beta \sqrt{\mathrm{Var}[\{S_i\}_{i\ge2}]},
\end{equation}
$O_1$ will be identified as an HFO, where $\alpha$ and $\beta$ are two
adjustable parameters, and $E[\cdot]$ and $\mathrm{Var}[\cdot]$ are the
expectation and variance functions. This process is repeated until no
on-intervals in the remaining sequence satisfy the criterion. Typically,
for one IMF, approximately $10\%$ of the on-intervals would be selected
as HFOs. Among the remaining HFOs, one can see that some are
so close to each other that it is reasonable to combine them. The
combinations are carried out wherever the gap $G$ between two
neighboring HFOs is small, i.e., when $G< g\cdot\min(T_{HFO_1},T_{HFO_2})$,
where $g$ is the parametric gap tolerance ratio and $T_{HFO}$ is the HFO
duration. An example of combining HFOs is shown in Fig.~\ref{fig:scheme}(b).

\begin{figure}
\centering
\includegraphics[width=\linewidth]{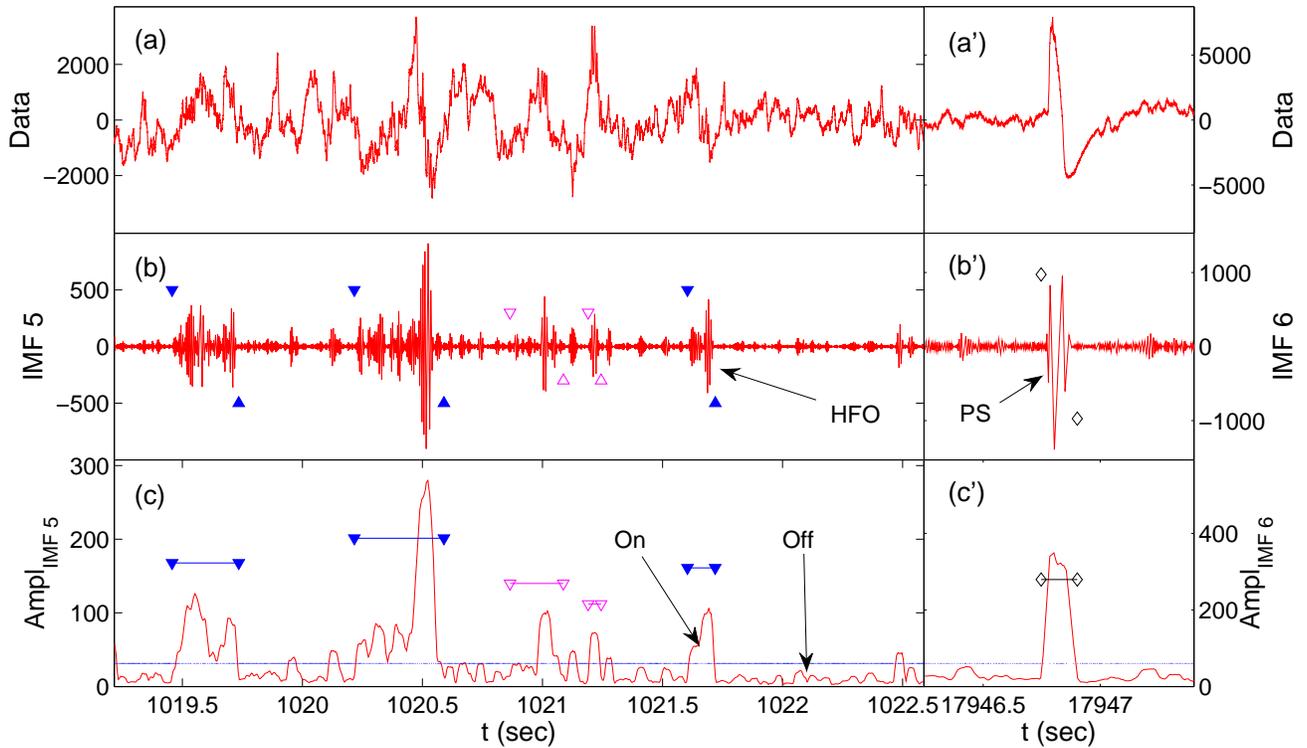}
\caption{\textbf{Example of successful HFO and PS detection.}
(a) Original EEG data plot of about 3 seconds. (b) IMF 5 plot with
solid blue triangles marking the ripples and open magenta triangles
marking the fast-ripples. (c) The amplitude of IMF 5. The horizontal
blue line is the threshold for separating on/off intervals of HFOs.
The threshold is calculated from the amplitude data segment of about one
hour. The computational parameters are $a_\mu=1$ and $a_\sigma=1$.
(a')-(c') The original data, IMF6, and its amplitude function, respectively.
The black diamonds mark the position of the population spikes.}
\label{fig:imf5hfo}
\end{figure}

HFOs in different frequency ranges are usually responsible for
different brain behaviors such as normal information processing and
spontaneous seizures. The third step of our procedure is to calculate
the frequencies of various HFOs, which is done by locating the
starting and ending times of an HFO, finding the number $n$ of oscillating
periods within it, and dividing by its duration: $f=n/T_{HFO}$.
When necessary, we combine overlapping/nearby HFOs, as shown in
Fig.~\ref{fig:scheme}(c). When various HFOs have been identified,
we can classify them into distinct frequency ranges: low-frequency
range ($<80$Hz), ripples [$80\sim200$Hz, between pairs of solid
blue triangles in Fig.~\ref{fig:scheme}(c)], and fast-ripples
[$> 200$ Hz, between pairs of open magenta triangles in
Fig.~\ref{fig:scheme}(c)]. HFOs of frequencies lower than $80$Hz
are identified as population spikes. An example of identifying
and classifying HFOs is shown in Fig.~\ref{fig:imf5hfo}.

\subsection{Statistical and scaling properties of HFOs}

As described, the on- and off-intervals associated with HFOs can be
determined by setting a threshold value $A_c$ in the amplitude function
of a given IMF. For a given HFO, an on-area can then be defined, which
is the area of the portion of the amplitude function above the threshold.
The most significant HFOs are those with the largest on-areas. The
on-intervals thus provide a base for characterizing the HFOs. Intuitively,
HFOs of various magnitude correspond to ``islands'' of various sizes above
the ``sea'' level defined by the threshold. To obtain a more complete
understanding of HFOs, it is insightful to examine the corresponding
``undersea'' dynamics (below the threshold). It is computationally feasible
to study the dynamical and statistical characteristics of the oscillations
below the ``sea'' level but only within a certain depth.

\begin{figure}
\centering
\includegraphics[width=\linewidth]{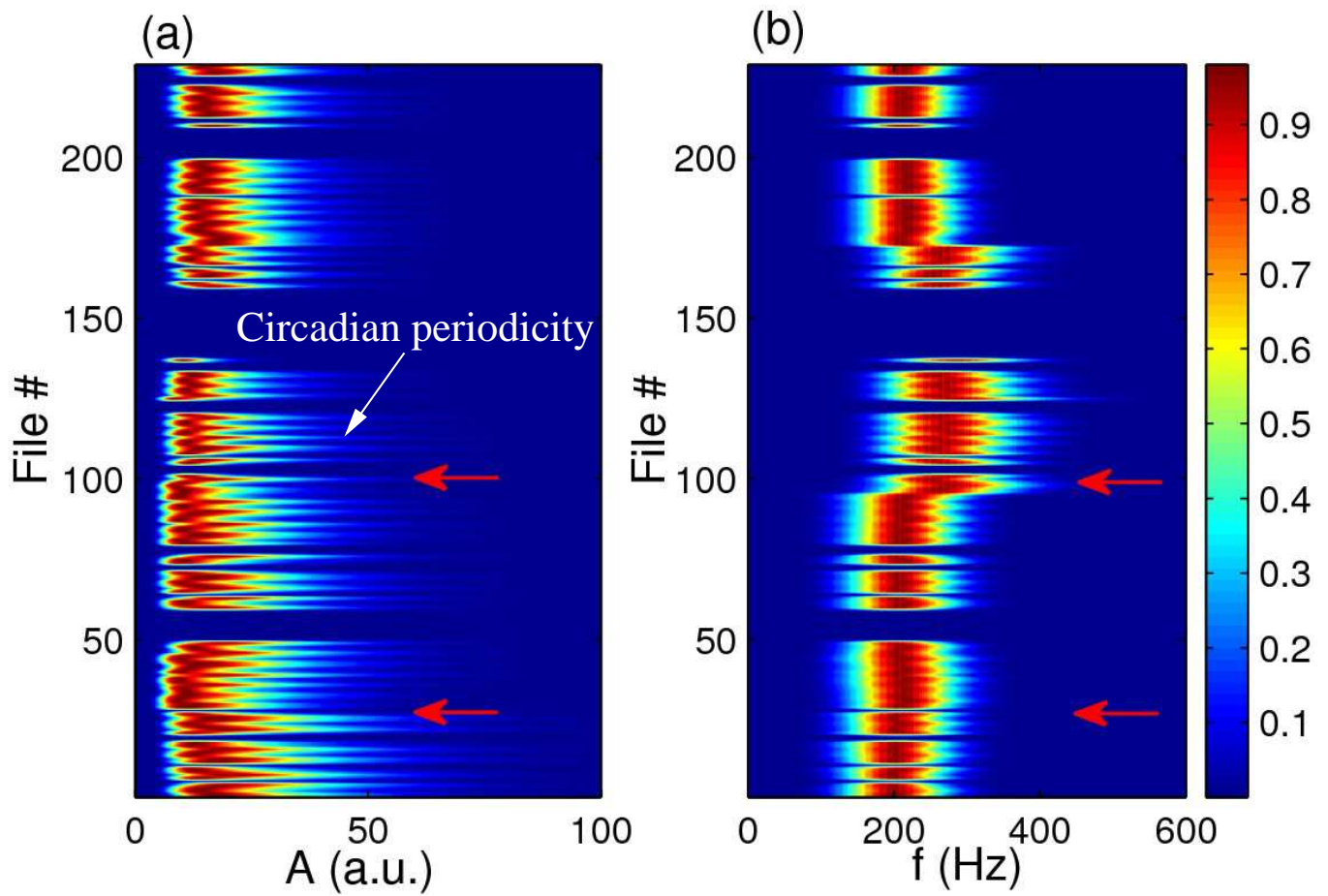}
\caption{\textbf{Typical EMD representation of massive rate EEF data.}
(a) Contour plot of normalized
distribution of amplitude $A$ (in arbitrary units) varying in
time of a particular EMD mode of interest (IMF5, around 200 Hz)
for channel 11 in CA1 of EEG recording of a rat over a 2-month
period. The all-blue region indicates corrupted files. Each file is a
7 hours recording at the sampling frequency $12$ kHz. Thus the
vertical axis ``File \#'' indicates time. The
distribution is calculated and then normalized by the maximum value
for each file. The rat underwent surgery between file 28 and file 29,
and the first seizure occurred in file 99, as indicated by the red
arrows. The comb-like structure indicates the circadian periodicity.
(b) Normalized distribution of the frequency $f$ of the mode.}
\label{fig:distribution}
\end{figure}

To illustrate our approach in a concrete way, we take the
example in Fig.~\ref{fig:imf5hfo} and focus on mode 5 because the
frequency of this mode lies in a suitable HFO frequency range.
From the contour plot of the distribution of amplitude
versus file number, Fig.~\ref{fig:distribution}, we see that there are
several regions of distinct properties. In particular, the EEG data are
relatively stable before the stimulation, say between files 28 and 29.
After the stimulation, the data changed characteristically, as can be seen
from the amplitude distribution plot in Fig.~\ref{fig:distribution}(a)
(indicated by the arrow between files 28 and 29). The first seizure
occurred in file 99, and the data are stable for files 29-99. There is
a relatively small change between files 51 and 52, as can be seen from
the amplitude plot in Fig.~\ref{fig:distribution}(a), when the rat
was actually moved from one cage to another. We have checked other modes
and also data from other channels and found a consistency in the specific
data segmentation as described. It is thus useful to study these different
segments separately: files 1-28 (before stimulation), files 29-51 (between
stimulation and first seizure), files 52-94 (preictal phase), files
100-172 (postictal phase but with recurrent seizures in files 105 and 125),
and files 175-223 (with recurrent seizures in files 192, 209, and 216).
To obtain stable statistics, we temporally disregard a few files
that are in the transitional regime. However, these files may be
important in providing possible hints about how the seizure (e.g., the
one occurred around file 99) is developed in terms of the transient
dynamics. The specific segmentation scheme of the EEG data is only for
one rat, but it is valid for all the channels.

\begin{figure}
\begin{center}
\includegraphics[width=\linewidth]{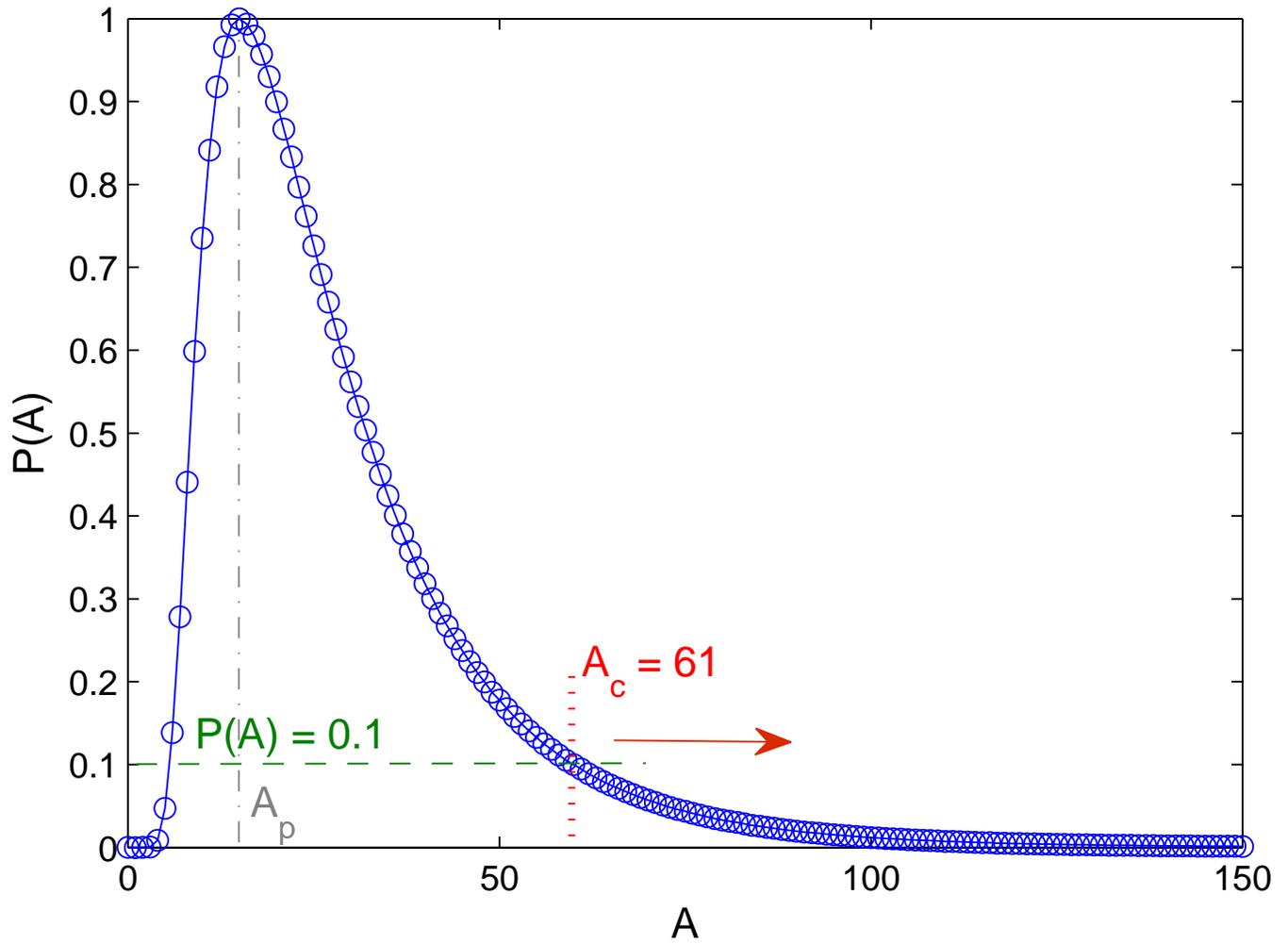}
\caption{\textbf{Determination of threshold $A_c$.} 
Normalized distribution $P(A)$ of the amplitude $A$ for files 1-28 for the 
same mode as in Fig.~\ref{fig:distribution}, where $A_p$ is the value of 
amplitude at the peak of the distribution. For example, by setting $P(A)=0.1$ 
and assuming that $A_c > A_p$, $A_c$ can be determined to be 61. If $P(A)$ 
takes a smaller value, then $A_c$ will be larger.}
\label{fig:distr}
\end{center}
\end{figure}

For different segments the signal amplitudes can have systematic differences,
and in certain cases the average amplitude can be, for example, twice larger in
one segment than in another. It is thus necessary to determine and set
different thresholds in different segments. 
To do this, first we calculate the normalized distribution of the amplitude 
$A$ in each segment. The distribution for files 1-28 is shown in 
Fig.~\ref{fig:distr}. Second, from the peak point defined as $P(A_p)=1$ 
(since $P$ is normalized to unity), we decrease $P$ with small steps, e.g., 
$P=0.98, 0.96, ..., 0.02$. For for each $p$ value, we determine the 
corresponding threshold $A_c$ ($>A_p$), as demonstrated in 
Fig.~\ref{fig:distr} for $P=0.1$, where the corresponding threshold 
is $A_c=61$. Third, for each value of $A_c$, we calculate all the 
on-intervals in this segment from the amplitude function.
The distributions of the durations $T$ of the on-intervals from different
segments are then compared, and various threshold values are set such that
the values of $P(A_c)$ are all identical across the segments.

\begin{figure}
\centering
\includegraphics[width=\linewidth]{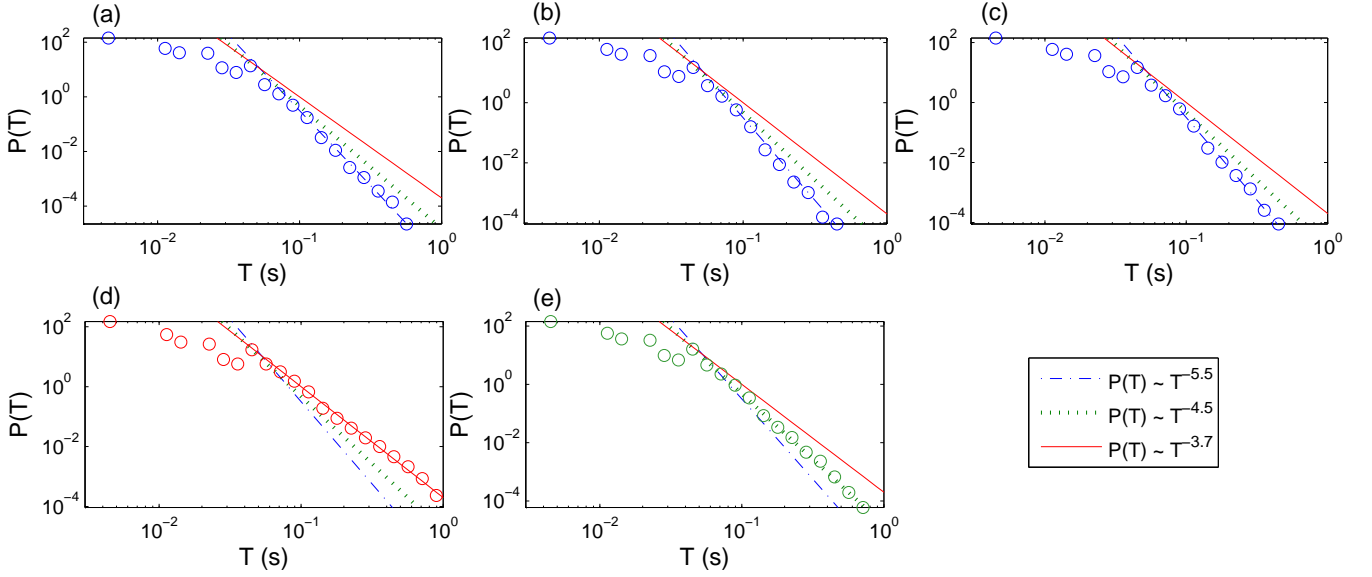}
\caption{\textbf{Statistical and scaling behaviors of HFOs.}
Distributions of on-interval $T$ of IMF 5 (channel 11,
Fig.~\ref{fig:distribution}): (a-e) for files 1-28, 29-51, 52-94,
100-172, 175-223 respectively. The numbers of on-intervals are
344310, 314698, 431674, 498947, and 510096 for (a-e), respectively.
An algebraic distribution is observed with different exponents for different
segments. The exponent for the solid, dotted, and 
dash-dotted lines are -3.7, -4.5, and -5.5, respectively.
The threshold $A_c$ is chosen such that $P(A_c)=0.02$ for
all the segments.}
\label{fig:onoff}
\end{figure}

Figure~\ref{fig:onoff} shows an example of the statistical distribution
of the on-intervals for all five segments as specified above. We
observe algebraic distributions. For each segment, the algebraic
scaling regime extends over at least one order of magnitude in the length of
the on-interval. The distributions for the three segments before the
first seizure have approximately the same algebraic scaling exponent, as
shown in Figs.~\ref{fig:onoff}(a-c), 
although the amplitude value varies continuously for these segments, 
as shown in Fig.~\ref{fig:distribution}. After the first seizure, the
exponent becomes smaller, indicating more on-intervals with longer
durations. For example, in Fig. \ref{fig:onoff}(d), for on-intervals
with $T \sim 0.4$ s, the probability is 100 times larger than those
before the seizure. For files 175-223, however, the exponent is somewhat
increased, signifying a decrease in the probability of longer on-intervals,
but it is still larger than the probability before the seizure.
We have systematically checked on-interval statistics for different
thresholds. The algebraic scaling and the qualitative difference
among the scaling exponents from different segments are robust with respect
to variations of the threshold in a certain range (e.g., $P(A_c)$ ranging
from 0.02 to 0.3).

Since many on-intervals (especially the long ones) correspond actually
to HFOs, our discovery of the algebraic scaling suggests that the HFOs
appear more active and sustaining associated with seizure activities,
which is consistent with previous observations~\cite{WGSH:2008,BES:2010,
CNHC:2010}. In general, an algebraic scaling indicates a hierarchical
organization in the underlying dynamics, which in our case, suggests
such an organization in the brain neuronal activities. For example, the
local synchrony among discrete neuron clusters may vary in hierarchical
scales. The fact that approximately the same algebraic scaling exponent
occurred before the seizure indicates that, after the stimulation, while
evolving toward epilepsy, the underlying dynamics behave the same as in the
normal brain. This could be due to the latency effect of the stimulation.
The development into epilepsy, however, occurs in a relatively short
period due to the cascading effect similar to that associated with
earthquakes~\cite{OFSML:2010}.

We have also checked other channels, which are so selected that they
belong to different (neural) correlation clusters. For some channels,
behaviors similar to those in Fig. \ref{fig:onoff} are observed, but
significant deviations occur for some other channels. This may be due to
that the HFO and the seizure onset zone are usually highly
localized \cite{CNHC:2010}. As a result, only within a proper
range of this zone can the HFOs be detected. Since the
distance between the neighboring channels is quite small (about
$0.25$ mm), the HFO and the underlying neuronal activity could be
revealed only in a small subset of channels.

\begin{figure}
\centering
\includegraphics[width=\linewidth]{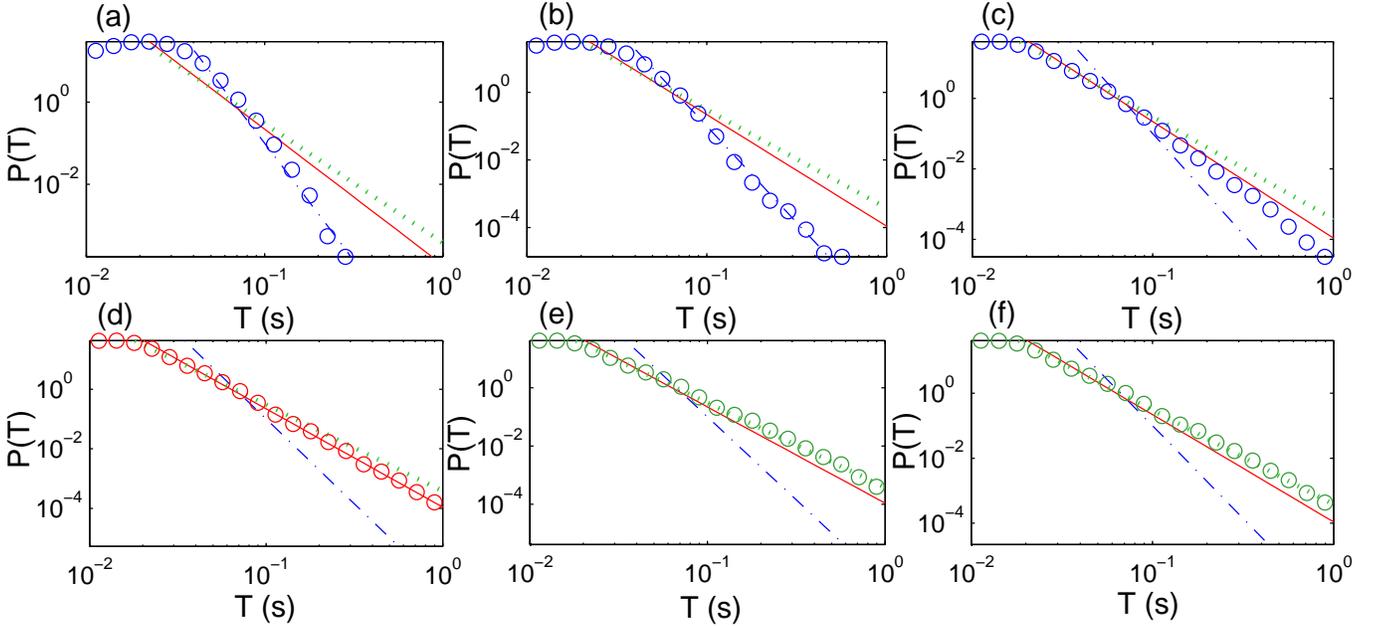}
\caption{
\textbf{Statistical and scaling behaviors of HFOs - more examples.}
Distributions of on-interval $T$ of IMF 5 (channel 6 of rat 9): (a-f) for 
files 1-19, 32-57, 63-72, 78-97, 98-118, and 119-149, corresponding to 
the pre-stimulation state, post-stimulation state, evolving towards seizure,
status epilepticus phase, epilepsy latent period, and spontaneous/recurrent 
seizure period, respectively. The numbers of on-intervals are 354499, 561669, 
300291, 458649, 293118, and 438919 for (a-f), respectively. An algebraic 
distribution is observed with different exponents for different segments, 
where the exponents are $-5.7$, $-3.3$, $-2.9$ for dash-dotted line, solid 
line, and dotted line, respectively. The criterion for choosing the threshold 
$A_c$ is the same as in Fig.~\ref{fig:onoff}}. 
\label{fig:onoff_more}
\end{figure}

We find that the algebraic scaling law for the on intervals of HFOs 
holds for all the animal models. Figures~\ref{fig:onoff_more}(a-f) show
more examples for a different rat. In particular, the scaling was calculated
for a specific channel for the pre-stimulation state, post-stimulation state, 
evolution towards seizure, the status epilepticus phase, 
the epilepsy latent period, and the spontaneous/recurrent seizure period, for 
panels (a-f), respectively. We see that, while the details of the scaling 
behaviors can be different for the distinct critical phases (e.g., during 
epileptogenesis the on-intervals with longer durations are dominant 
after the first seizure), the algebraic nature of the scaling law is robust.

\section{Discussion}

Seizure prediction, early recognition, and blockage of seizures are
considered by the membership of the American Epilepsy Society (AES) as
the first research priority listed among fifteen. To achieve these goals
a good understanding of the origin, mechanism, and dynamics of
seizures is necessary. At present the only accessible avenue to probe
the origin of epileptic seizures is multiple-channel EEG or ECoG data.
Continuous improvement in the experimental methodology has made such data
highly reliable and generally of high quality. A challenge is that the amount
of EEG or ECoG data is {\em massive}.  An issue of paramount importance
and significant interest is to {\em extract knowledge about epilepsy
from data}.

We have developed a method to detect, characterize, and
analyze the dynamical behavior of HFOs from a massive database of
extensive EEG recordings of a number epileptic rats over two months.
We first devise a general and efficient procedure
for pre-processing the massive database to exclude erroneous data
segments and statistical anomalies. (The procedure should also be
applicable to massive data sets of other sorts, such as large-scale
sensor data or seismic data collections.) We then articulate a procedure
based on separating the time scales, the EMD
method~\cite{Huangetal:1998,YL:1997}, to detect
HFOs. Due to its built-in ability to pick out
from a complex, seemingly random time series a number of dominant
components of distinct time scales, the method has proven to be
especially effective for the detection of HFOs. We finally perform a
statistical analysis and find evidence for a striking
phenomenon: the occurrences of HFOs appear in an on-off intermittent
manner and the time intervals that they last exhibit an algebraic
scaling.

The methods and results in this paper can potentially be extended to
other fields. For example, a previous work~\cite{OFSML:2010} indicated
that seizures can be regarded as ``quakes of the brain.'' It is then
conceivable that the idea, method, and algorithms developed here can
be extended to big seismic database for detecting anomalous oscillating
signals (similar to HFOs) preceding the actual occurrence of earthquakes.

Big data sets arise not only in biomedicine,
but also in other fields of science and engineering. For example, in
civil engineering, large sensor arrays are often employed to monitor
the temperature, humidity, and energy flows in large, complex infrastructure
systems in a continuous-time fashion. In such an application the underlying
system is in general nonstationary and nonlinear, and to detect behaviors
that deviate from the norm or the expected is of considerable interest.
Another example is earthquake data. A previous work~\cite{OFSML:2010} 
indicated that seizures can be regarded as ``quakes of the brain.'' It is 
then conceivable that the idea, method, and algorithms developed in this
paper can be extended to big seismic database for detecting anomalous 
oscillating signals (similar to HFOs) preceding the actual occurrence of 
earthquakes. In general, the EMD/Hilbert transform based methodology 
demonstrated in this paper has broad applications because it is specifically 
designed~\cite{Huangetal:1998} to deal with nonlinear and nonstationary 
systems. Due to its built-in ability to obtain from a complex, seemingly 
random time series a number of dominant components of distinct time scales, 
the method is anticipated to be especially effective for anomaly detection.
A challenge is to develop mathematically justified, computationally 
reasonable and automated procedures to detect anomalies, from big data sets, 
which has been addressed in this paper through the detection of HFOs from 
massive EEG data with multiple animal models.

We now discuss a number of issues that may warrant future efforts.

First, there is room to improve the EMD-based algorithms developed
in this paper, leading possibly to a fully automated method for
detecting HFOs and population spikes from massive data for all
distinct epileptic stages including pre-stimulation, pre-seizure and
recurrent seizures. This will provide a base for probing into the
emergence and evolution of HFOs through detailed analyses using methods
from nonlinear dynamics, statistics, and statistical physics. Special
features associated with different types of brain activities can
be identified, with the grand goal to exploit the predictive power
of HFOs for epileptic seizures. Many questions, which are previously
unimaginable, can be asked. For example, does a general class of
HFOs exist, regardless of the specific brain activities? What types
of HFOs are especially related to seizures? How localized are they
around the regions of seizure onset? Are there systematic and
characteristic changes in the HFOs during the ictal phase? What
types of HFOs are associated with recurrent seizures and is there
any relation to the latency period? Answers to those questions and
many more will provide a comprehensive picture for the dynamical
role of HFOs in epileptogenesis.

Second, with our optimized EMD-based method, HFOs and population spikes
can be detected reliably for all the channels. The issue of spatiotemporal
evolution of these dynamical events in the brain can then be addressed.
For example, we have observed that certain population spikes appear
only in some channels at a given time, i.e., they may be highly
localized in space. Some HFOs can, however, occur in many channels
simultaneously. A possible reason for the dispersive character of
the HFOs is that the distance between neighboring recording sites
is about 0.25mm, which can be within the propagating range of the
underlying neuronal activities that cause the pathological HFOs.
The available database from a dense electrode grid thus provides a
useful probe to study the propagation of neuronal synchronous activities.
For example, by examining the exact timings of HFOs and population spikes
and the occurrence of the epileptic seizure in different EEG channels, the
propagating pattern of these dynamical events can be determined and,
consequently, the sources triggering these events may be identified. A mapping
between the epileptic dynamics and activities in different brain regions
can be made and the temporal evolution of the mapping can be studied.
It would also be useful to examine the correlation patterns of the
distinct dynamical events and compare them with those of the background
neuronal activities. All these have the potential to provide deeper
insights into the origin of epileptogenesis.

Third, in this paper we focused on stable states of
the brain, which are the states that last for relatively long
periods of time. The transient behaviors were neglected.
It would be interesting to study the nonlinear and complex
transient dynamics~\cite{LT:book} associated with epileptogenesis
and brain behaviors in general.

Fourth, there were previous studies of exploiting nonlinear dynamics
for analyzing epileptic seizures (e.g., Refs.~\cite{OHLF:2001,
LOHF:2002,LHFO:2003,LHFO:2004,HOFAL:2005,LFO:2006,LFOH:2007}),
and on-off intermittency is an ubiquitous phenomenon in nonlinear
dynamical systems~\cite{FY:1985,FY:1986,FIIY:1986,PST:1993,HPH:1994,
PHH:1994,HPHHL:1994,SO:1994,LG:1995,VAOS:1995,AS:1996,Lai:1996a,
Lai:1996b,VAOS:1996,MALK:2001,RC:2007}. It is known in nonlinear
dynamics that on-off intermittency can be controlled by applying small but
deliberately chosen perturbations to the system~\cite{NHL:1996}. 
The power-law statistics of the on-intervals associated with HFOs indicate 
the emergence of a hierarchical organization in the neuronal activities. 
However, it is difficult to determine uniquely the underlying dynamical 
mechanism that is distinct from the well documented mechanism for on-off 
intermittency. Nonetheless, there are common features between the 
intermittent behaviors of HFOs uncovered in this paper and those of on-off 
intermittency, such as scaling properties. If a specific type of HFO can 
be correlated with the occurrence of seizures and if the underlying 
dynamics bear similarity to those of generic on-off 
intermittency, it may be possible to investigate controlling HFO dynamics 
based on previous works on controlling on-off intermittency.
In particular, can the on-off intermittent dynamics of the epileptic
HFOs be controlled by using, say, small and infrequent brain stimuli to
delay or even eliminate seizures? The results in this paper provide a
base for developing computational and experimental schemes to test this
control idea.

\section{Methods}

{\em All experiments on live vertebrates (rats) were performed in accordance
with the guidelines and regulations of National Institute of Health and
University of Florida. All experimental protocols were approved by
University of Florida.}

\subsection{Setting of experimental data collection}

Experiments were performed on five, two-month old, male Sprague Dawley
rats, each weighing between 210 and 265 g. The rats were induced into the
state of complete anesthesia by subcutaneous injection of 10 mg/kg
(0.1 mL by volume) of xylazine and maintained in the anesthetized state
using isoflorane (1.5) administered through inhalation by a precision
vaporizer. For each rat, the top of its head was shaved and
chemically sterilized with iodine and alcohol. The skull was exposed
by a mid-sagittal incision. In three of the five rats, a bipolar,
twisted, Teflon-coated, stainless steel electrode (330$\mu$m) was
implanted in the right posterior ventral hippocampus (5.3mm caudal
to bregma, 4.9 mm right of midline suture and at a depth of 5 mm
from the dura) for stimulating the rat into status epilepticus.
The remaining two rats are for control. In
all the five rats a 16 microwire (50 $\mu$m, TDT Technologies,
Alachua, FL) electrode array comprising of two rows separated by 500
$\mu$m with electrode spacing of 250 $\mu$m, was implanted to the
left of midline suture horizontally in the CA1-CA2 and
dentate gyrus of the hippocampus. The furthest left microwire was
4.4 mm caudal to bregma, 4.6 mm left of midline suture and at a
depth of 3.1 mm from the dura. A second microwire array of 16
electrodes was implanted to the right of midline suture in a
diagonal fashion. The furthest right microwire was 3.2 mm caudal to
bregma, 2.2 mm to the right of midline suture. The closest right
microwire was 5.2 mm caudal to bregma, 1.7 mm to the right of
midline suture and at a depth of 3.1 mm from the dura. Finally four
0.8 mm stainless steel screws were placed in the skull to anchor the
microwire electrode array: two screws were AP 2 mm to the bregma and
bilaterally 2 mm and served as the ground electrodes while two
screws were AP -2 mm to the lambdoidal suture and bilaterally 2 mm
and served as the reference electrodes. The entire surgical area was
then closed and secured with cranioplast cement. Following surgery
the rats were allowed to recover for a week.

\subsection{Data acquisition and file structures}

Electrophysiological recordings were conducted by hooking each rat onto
a 32-channel commutator, the output of which was fed
into the recording system comprised of two 16-channel pre-amps, which
digitized the incoming signal with a 16 bit A/D converter at a sampling
rate of 12kHz ($\sim 12207$ Hz). The digitized signal was then sent over
a fiber optic cable to a Pentusa RX-5 data acquisition board (Tucker
Davis Technologies), where the signals were bandpass-filtered between
1.5 Hz and 7.5 kHz. The digital stream of data was
stored for later processing. For each channel, the data were recorded and
saved as $16$-bit signed integer binary files, each of size $600-700$
megabytes ($\sim 7.5$ hours recording time). Thus, for each rat, there were
$32$ channels, each of which has between 153 and 317 files depending on the
recording durations. Each binary file was assigned a {\em rat} number, a {\em
channel} number, and a {\em file} number.
The data were recorded over 2 months for most of the rats, including
pre-stimulation stage, pre-seizure stage, status epilepticus phase,
epilepsy latent period, and spontaneous/recurrent seizure period.

The sampling rate of the data was relatively high, making it possible to
analyze high frequency, short duration dynamical events in the brain
such as HFOs and population spikes. The typical duration of an HFO
is about 100 ms and its characteristic spontaneous frequency can be as high as
several hundred Hertz. In such a case, our data will have 40 points
for a single oscillation period, which is generally enough for various
analyses of HFOs. The extensive database provides us with
a platform to compare the high frequency events in different stages
in the evolution of epileptic seizures and to systematically investigate
the dynamics of epileptogenesis.

\subsection{Empirical Mode Decomposition (EMD) of EEG data}
 
EMD is specifically designed to deal with nonlinear and non-stationary
data sets. In particular, EMD decomposes the signal into a series of
intrinsic mode functions (IMFs), as follows. For a given signal, EMD
determines the local maxima and local minima, and connect them with
cubic splines to form an IMF. One then subtracts the IMF from the
original signal, and repeats the process to get the second IMF, and so on.
The procedure is repeated until the remaining signal becomes monotonic. The
IMFs are orthogonal to each other (at least locally) and their sum
restores the original data. Thus, effectively, the original signal
has been decomposed into the IMFs, each in a distinct frequency range,
whose non-stationary amplitude and frequency information is well preserved.

Ideally, for a given EEG data, the EMD method returns a set of IMFs in
separate frequency ranges. Practically, since each data file is too
large to be processed computationally, we need to divide the data into
small segments so that each segment
can be computed within the memory of our computers. To deal with the
boundary effect properly, for each data segment, we include an extra
but much smaller subset of data points on both ends of the segment, which
are from neighboring segments. These are the corresponding boundary
sets. After performing the EMD calculations,
only the IMFs within the original data segment are kept, while those
associated with the boundary sets are discarded. For a given
data segment, the resulting IMFs usually depend on the
choices of the sizes of the segment and the boundary sets.
In particular, the larger the boundary sets, the more accurate
the IMFs, but the amount of the computation will also increase.
A systematic test on varying sizes of the boundary
sets indicates that the choice of 0.5-second duration (corresponding
to 6103 data points at the recording sampling frequency) for each
boundary set yields accurate IMFs with tolerable extra computation
load. The limited computational power also stipulates that the
size of each segment itself cannot be too large. Our systematic
test gave 5 seconds (about 61035 data points) as the optimal duration
for balancing the computation time and the reliability of the results.
We use the code developed by Rilling {\it et al.}~\cite{EMDcode} to
perform the EMD calculations by modifying the original C-Matlab interface
to C-codes.

\begin{figure}
\centering
\includegraphics[width=\linewidth]{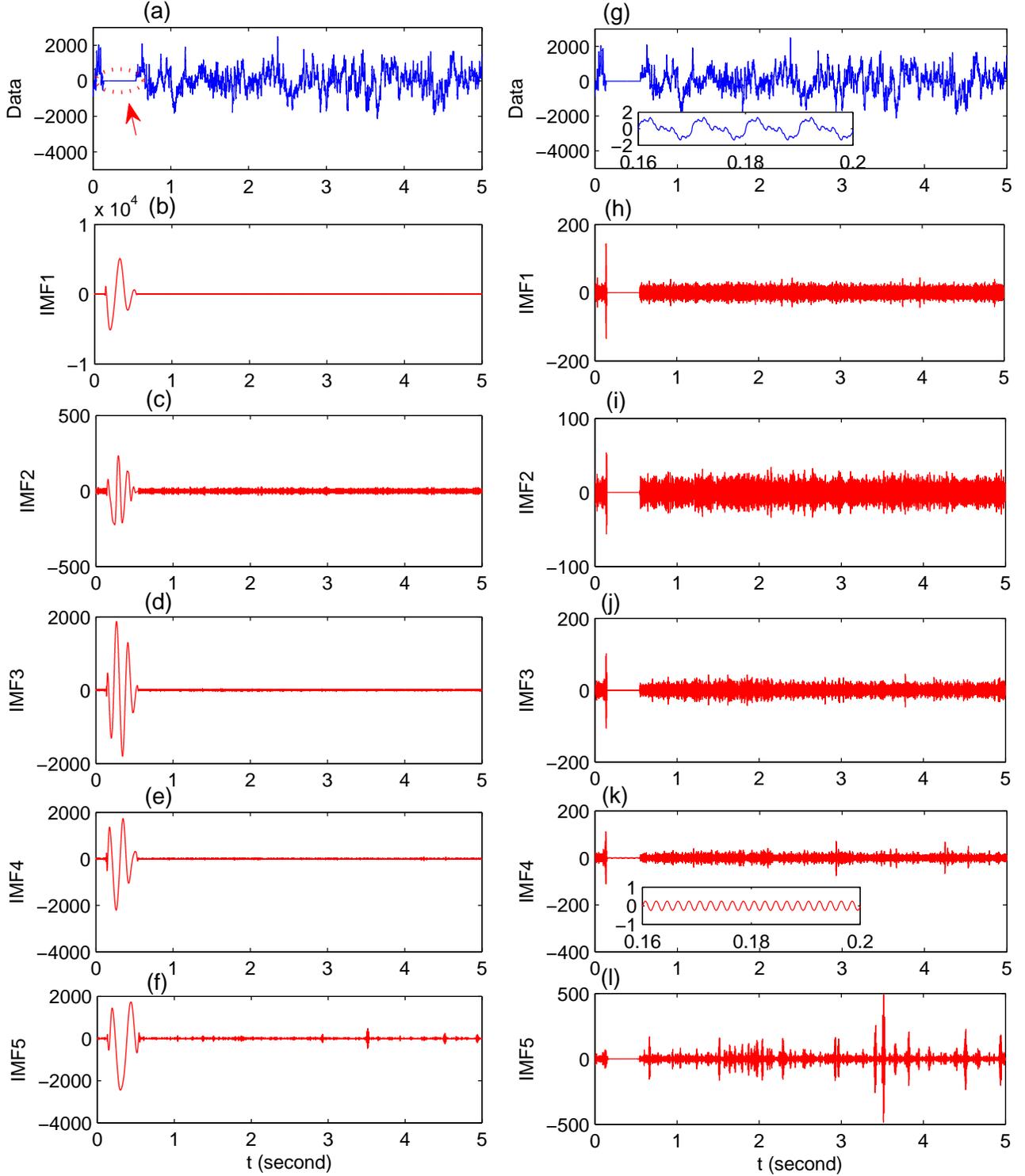}
\caption{\textbf{A demonstration of adding small oscillations 
in EMD computation to eliminate edge effect.} (a) A 5 second segment of 
data with about 0.4 second zeros, as indicated by the dotted circle. 
(b-f) The first 5 intrinsic mode functions directly calculated from the data
in (a). (g) Data (a) with added small oscillations on the scale of unity 
(see text), which is almost invisible from the figure. (h-l) The 
first 5 intrinsic mode functions calculated from the data in (g).
Insets of (g) and (k) show magnification of the zero region. Note
that the scale of the $y$ label is much larger in (b-f) than those
in (h-l). The anomalies appeared in (b-f) are effectively removed by
the simple method of adding small oscillations at the boundaries of 
the data segment.}
\label{fig:EMDtrick}
\end{figure}

Another practical issue is that the data may contain some
discontinuities. In such a case, the EMD program may diverge
or have abnormally large values [Fig.~\ref{fig:EMDtrick}(a-f)]. 
A remedy is to add a small perturbation in the original signal prior to
the EMD calculations. However, due to the difference in
the frequency ranges in which the various IMFs lie, small time-varying
perturbation signals of frequencies in these distinct ranges
are needed. For each frequency range, the amplitude of the perturbation
needs to be orders of magnitude smaller than that of the corresponding
IMF. For example, if for a normal EEG data segment (denoted by $y[i]$),
there are six IMFs and their frequencies are about 5 kHz, 2 kHz, 1 kHz,
500 Hz, 200 Hz, and 100 Hz, respectively (for IMFs 1-6), we will need to
add the following small sinusoidal signals:
$y[i]=y[i] + 0.9    \times \sin{(2\pi i/(12207/100))}
                + 0.5    \times \sin{(2\pi i/(12207/200 ))}
                + 0.25   \times \sin{(2\pi i/(12207/500 ))}
                + 0.125  \times \sin{(2\pi i/(12207/1000))}
                + 0.0625 \times \sin{(2\pi i/(12207/2000))}
                + 0.03   \times \sin{(2\pi i/(12207/5000))}$,
where the amplitudes of the IMFs are typically larger than 10.
The perturbation signals thus will not have any practical influences on
the IMF results for normal data. However, when there is a discontinuity
with a linear relaxation in time, the corresponding IMFs will contain the
added small sinusoidal oscillations instead of generating divergence or large
anomalies [Fig.~\ref{fig:EMDtrick}(g-l)]. 
In addition, when the original data is contaminated by a small segment
of zeros, without adding the small oscillations, the resulting IMFs
will oscillate wildly in this region with amplitudes orders of magnitude
larger than those of the normal data sets [Fig.~\ref{fig:EMDtrick}(a-f)].
This is because, for obtaining each
IMF, EMD looks for the local maxima and local minima and then
approximate the data with cubic spline connecting the maxima or minima.
When a segment of zeros is encountered, there are no local maxima or minima
so that the EMD extrapolates with cubic spline using the maxima or minima
outside this region. For the first IMF, since the frequency is the highest
(about 5 kHz), even a zero segment of about 0.1 second
would correspond to about 500 maxima or minima. Thus the extrapolation
will generate extremely large, artificial oscillations. The
remainder obtained by subtracting IMF 1 from the original data will compensate
the large oscillations in IMF 1, but they will propagate to subsequent IMFs.
The conclusion is that, adding the small sinusoidal perturbing signals
causes essentially no difference in the original signal (about 1 part in
1000), but the artificial anomalies can be effectively eliminated.

\section*{Research Ethics}

Human datasets were not used.

\section*{Animal ethics}

Experiments were performed on 2-month old male Sprague Dawley rats.  This study was conducted in accordance with Federal and University of Florida Institutional Animal Care and Use Committee policies regarding the ethical use of animals in research.  (IACUC protocol D710).

\section*{Permission to carry out fieldwork}

No fieldwork was performed

\section*{Data Availability}

All data used in the study have been uploaded onto Google Drive and are 
publicly available. The link is
https://drive.google.com/drive/folders/0B7S5nQOU$_{-}$nMIYkEyYmJvTGpNOVU?usp=sharing. 

\section{Competing Interests}

We have no competing interests.

\section*{Authors' Contributions}

YCL, LH, PRC, WLD and MLS conceived and designed the research.  The data was acquired in PRC’s lab.  LH and XN developed the computational method and performed simulations.  All analyzed data.  YCL and LH drafted the manuscript.  All authors gave final approval for publication.

\section*{Funding}

The National Institutes of Biomedical Imaging and Bioengineering (NIBIB) through Collaborative Research in Computational Neuroscience (CRCNS) Grant Numbers R01 EB004752 and EB007082, the Wilder Center of Excellence for Epilepsy Research, and the Children’s Miracle Network supported this research. This work was also supported by the US Army Research Office under Grant No. W911NF-14-1-0504. LH was supported by NNSF of China under Grants No.~11135001, No.~11375074, and No.~11422541. Y.C.L. would like to acknowledge support from the Vannevar Bush
Faculty Fellowship program sponsored by the Basic Research Office of
the Assistant Secretary of Defense for Research and Engineering and
funded by the Office of Naval Research through Grant No.~N00014-16-1-2828.


\clearpage

\end{document}